\documentclass[english]{article}
\usepackage[T1]{fontenc}
\usepackage[latin9]{inputenc}
\usepackage{geometry}
\usepackage{color}
\usepackage{array}
\usepackage{float}
\usepackage{amsmath}
\usepackage{graphicx}
\usepackage{amssymb}

\makeatletter

\providecommand{\tabularnewline}{\\}

\makeatother

\usepackage{babel}

\begin{document}

\title{Relation between squeezing of vacuum fluctuations, quantum entanglement
and sub-shot noise in Raman scattering}

\maketitle
\begin{center}
Anirban Pathak$^{a,b}$%
\footnote{email: anirban.pathak@gmail.com%
}, Jaromir ${\rm K}\breve{{\rm r}}{\rm epelka}^{b}$ and Jan ${\rm \check{Perina}}$$^{b,c}$ 
\par\end{center}

\begin{center}
$^{a}$Jaypee Institute of Information Technology, A-10, Sector-62,
Noida, UP-201307, India 
\par\end{center}

\begin{center}
$^{b}$RCPTM, Joint Laboratory of Optics of Palacky University and
Institute of Physics of Academy of Science of the Czech Republic,
Faculty of Science, Palacky University, 17. listopadu 12, 771 46 Olomouc,
Czech Republic
\par\end{center}

\begin{center}
$^{c}$Department of Optics, Palacky University, 17. listopadu 12,
771 46 Olomouc, Czech Republic
\par\end{center}
\begin{abstract}
A completely quantum description of Raman process is used to investigate
the nonclassical properties of the modes in the stimulated, spontaneous
and partially spontaneous Raman process. Both coherent scattering
(where all the initial modes are coherent) and chaotic scattering
(where initial phonon mode is chaotic and all the other modes are
coherent) are studied. Nonclassical character of Raman process is
observed by means of intermodal entanglement, single mode and intermodal
squeezing of vacuum fluctuations, sub-shot noise and wave variances.
Joint photon-phonon number and integrated-intensity distributions
are then used to illustrate the observed nonclassicalities. Conditional
and difference number distributions are also provided to illustrate
the nonclassical character. The mutual relation between the obtained
nonclassicalities and their variations dependent on phases, rescalled
time and ratio of coupling constants are also reported. 
\end{abstract}

\section{Introduction}

Quantum statistics of Raman scattering were discussed from various
points of view in a number of papers (see \cite{B Sen-Sub-shot-noise,Bsen-2}
and references therein, Section 10.4 of \cite{The book of Prof. Perina}
and \cite{modern nonlinear optics} for reviews). In this paper we
describe the Raman scattering process with a completely quantum mechanical
Hamiltonian. The model is capable to include the stimulated, spontaneous
and partially spontaneous Raman process. We use the second order short-time
approximation for solution of the Heisenberg equation of motion corresponding
to this Hamiltonian. The solution is then used to relate the nonclassical
properties of photons and phonons in these processes (i.e. stimulated,
spontaneous and partially spontaneous Raman process). To be precise,
nonclassical characteristics of photon and phonon modes generated
in these processes are exhibited through a number of properties, including
squeezing of vacuum fluctuations, quantum entanglement of modes, sub-shot
noise and wave variances. Further, joint photon-phonon number, integrated-intensity
distributions, conditional and difference number distributions are
also found useful to illustrate the observed nonclassicalities. 

Remaining part of the paper is arranged as follows. In Section \ref{sec:The-model-Hamiltonian}
we briefly describe the fully quantum Hamiltonian of Raman process
and several criteria of nonclassicalities that are used in the present
study. In Section \ref{sec:The-model-Hamiltonian} we have briefly
described the model Hamiltonian that provides a completely quantum
description of the Raman scattering process and also describe a normal-ordered
characteristic function in the Gaussian form for the same system.
Criteria for testing of nonclassicalities are also introduced. In
Section \ref{sec:Phonon-mode-is-cohrent} we investigate nonclassical
character of Raman process for coherent scattering by means of intermodal
entanglement, single mode and intermodal squeezing of vacuum fluctuations,
sub-shot noise and wave variances. In Section \ref{sec:Phonon-mode-is-chaotic}
the same nonclassical characteristics are investigated for the chaotic
scattering. The observed nonclassicalities are further illustrated
through joint photon-phonon number and wave distribution in Section
\ref{sec:Joint-photon-phonon-number}. In Section \ref{sec:Difference-and-conditional}
we study difference and conditional number distributions associated
with the Raman process. Finally, Section \ref{sec:Conclusion} is
dedicated to conclusion.

\section{The model Hamiltonian and the criteria of nonclassicality\label{sec:The-model-Hamiltonian}}

A fully quantum description of the Raman scattering process can be
provided by the following Hamiltonian \cite{B Sen-Sub-shot-noise,Walls}:
\begin{equation}
H=\sum_{j=L,S,A,V}\hbar\omega_{j}a_{j}^{\dagger}a_{j}-\left(\hbar ga_{L}a_{S}^{\dagger}a_{V}^{\dagger}+\hbar\chi^{*}a_{L}a_{V}a_{A}^{\dagger}+{\rm h.c.}\right),\label{eq:hamiltonian}\end{equation}
where ${\rm h.c.}$ stands for Hermitian conjugate and the subscripts
$L,S,A$ and $V$ correspond to pump (laser), Stokes, anti-Stokes
and vibration (phonon) modes respectively, $\omega_{j},\, a_{j}$
and $a_{j}^{\dagger}$ are frequency, annihilation operator and creation
operator in the $j$ th mode, $g$ and $\chi$ are the Stokes and
anti-Stokes coupling constants. Using the Hamiltonian (\ref{eq:hamiltonian})
we can construct a set of Heisenberg equations and solve them in short-time
approximation. A second order short-time approximated solution was
already reported \cite{The book of Prof. Perina}. It is interesting
to note that using the short-time approximated solution, we can obtain
a normal-ordered characteristic function in the Gaussian form. Such
a characteristic function can completely characterize the Raman process,
and it can be analytically expressed as \cite{The book of Prof. Perina}
\begin{equation}
\begin{array}{lcl}
C_{N}(\beta_{L},\beta_{S},\beta_{A},\beta_{V},t) & = & \left\langle \exp\left\{ \sum_{j=L,S,A,V}\right.\left[-B_{j}(t)|\beta_{j}|^{2}\right.\right.\\
 & + & \left.\left(\frac{1}{2}C_{j}^{*}(t)\beta_{j}^{2}+{\rm c.c.}\right)+\beta_{j}\xi_{j}^{*}(t)-\beta_{j}^{*}\xi_{j}(t)\right]+\\
 & + & \left.\left.\sum_{j<k}\left(D_{jk}(t)\beta_{j}^{*}\beta_{k}^{*}+\bar{D}_{jk}(t)\beta_{j}\beta_{k}^{*}+{\rm c.c.}\right)\right\} \right\rangle ,\end{array}\label{eq:characteristic function}\end{equation}
 where ${\rm c.c.}$ stands for complex conjugate terms, the set $(L,\, S,\, A,\, V)$
is assumed to be ordered and \begin{equation}
\begin{array}{lcl}
B_{L}(t) & = & |\chi|^{2}t^{2}|\xi_{A}|^{2},\\
B_{S}(t) & = & |g|^{2}t^{2}|\xi_{L}|^{2},\\
B_{A}(t) & = & 0,\\
B_{V}(t) & = & |g|^{2}t^{2}|\xi_{L}|^{2}+|\chi|^{2}t^{2}|\xi_{A}|^{2},\\
C_{L}(t) & = & -g^{*}\chi t^{2}\xi_{S}\xi_{A}\exp(-2i\omega_{L}t),\\
C_{S}(t) & = & 0,\\
C_{A}(t) & = & 0,\\
C_{V}(t) & = & -g\chi t^{2}\xi_{S}^{*}\xi_{A}\exp(-2i\omega_{V}t),\\
D_{LS}(t) & = & -\frac{1}{2}|g|^{2}t^{2}\xi_{L}\xi_{S}\exp\left[-i\left(\omega_{L}+\omega_{S}\right)t\right],\\
D_{LA}(t) & = & -\frac{1}{2}|\chi|^{2}t^{2}\xi_{L}\xi_{A}\exp\left[-i\left(\omega_{L}+\omega_{A}\right)t\right],\\
D_{LV}(t) & = & \left[i\chi t\xi_{A}-\frac{1}{2}\left(|g|^{2}+|\chi|^{2}\right)t^{2}\xi_{L}\xi_{V}\right]\exp\left[-i\left(\omega_{L}+\omega_{V}\right)t\right],\\
D_{SA}(t) & = & -\frac{1}{2}g\chi^{*}t^{2}\xi_{L}^{2}\exp\left[-i\left(\omega_{S}+\omega_{A}\right)t\right],\\
D_{SV}(t) & = & \left(igt\xi_{L}-\frac{1}{2}|g|^{2}t^{2}\xi_{S}\xi_{V}-g\chi t^{2}\xi_{A}\xi_{V}^{*}\right)\exp\left[-i\left(\omega_{S}+\omega_{V}\right)t\right],\\
D_{AV}(t) & = & -\frac{1}{2}|\chi|^{2}t^{2}\xi_{A}\xi_{V}\exp\left[-i\left(\omega_{A}+\omega_{V}\right)t\right],\\
\bar{D}_{LS}(t) & = & -\frac{1}{2}g\chi^{*}t^{2}\xi_{L}\xi_{A}^{*}\exp\left[i\left(\omega_{L}-\omega_{S}\right)t\right],\end{array}\label{eq:BCD and Dbar}\end{equation}
 all other $\bar{D}_{jk}=0;$ $\xi_{j},\, j=L,S,A,V$ are initial
coherent complex amplitudes. As the above characteristic function
is Gaussian consequently (\ref{eq:BCD and Dbar}) can be used to obtain
normal fluctuation quantities (variances) $\left\langle \left(\Delta W_{j}\right)^{2}\right\rangle _{N}$
and $\langle\Delta W_{j}\Delta W_{k}\rangle_{N}$, which are defined
as \cite{The book of Prof. Perina}: \begin{equation}
\begin{array}{lcl}
\left\langle \left(\Delta W_{j}\right)^{2}\right\rangle _{N} & = & \left\langle a_{j}^{\dagger2}(t)a_{j}^{2}(t)\right\rangle -\left\langle a_{j}(t)a_{j}(t)\right\rangle ^{2}\\
 & = & \left\langle B_{j}^{2}+|C_{j}|^{2}+2B_{j}|\xi_{j}(t)|^{2}+\left[C_{j}\xi_{j}^{*2}(t)+{\rm c.c.}\right]\right\rangle \end{array}\label{eq:variance}\end{equation}
 and \begin{equation}
\begin{array}{lcl}
\langle\Delta W_{j}\Delta W_{k}\rangle_{N} & = & \left\langle a_{j}^{\dagger}(t)a_{k}^{\dagger}(t)a_{j}(t)a_{k}(t)\right\rangle -\left\langle a_{j}^{\dagger}(t)a_{j}(t)\right\rangle \left\langle a_{k}^{\dagger}(t)a_{k}(t)\right\rangle \\
 & = & \left\langle |D_{jk}|^{2}+|D_{jk}|^{2}+\left[D_{jk}\xi_{j}^{*}(t)\xi_{k}^{*}(t)-\bar{D}_{jk}\xi_{j}(t)\xi_{k}^{*}(t)+{\rm c.c.}\right]\right\rangle .\end{array}\label{eq:cross variance}\end{equation}
Brackets on the right-hand side in (\ref{eq:characteristic function},
\ref{eq:variance} and \ref{eq:cross variance}) mean an average over
the initial amplitudes. Equations (\ref{eq:BCD and Dbar})-(\ref{eq:cross variance})
provide us with sufficient mathematical framework required for the
study of the nonclassical character of stimulated and spontaneous
Raman process. This is so because the criteria for various nonclassical
phenomena can be conveniently expressed in terms of the quantities
described in (\ref{eq:BCD and Dbar})-(\ref{eq:cross variance}).
For example, we may note the criteria for principle squeezing of vacuum
fluctuations in single mode ($j)$ and compound mode ($ij),$ which
are \cite{The book of Prof. Perina}: \begin{equation}
\lambda_{j}=1+2(B_{j}-|C_{j}|)<1\label{eq:single-mode squuezing}\end{equation}
 and \begin{equation}
\lambda_{ij}=1+B_{i}+B_{j}-2{\rm Re}\bar{D}_{ij}-|C_{i}+C_{j}+2D_{ij}|<1\label{eq:squeezing compund mode}\end{equation}
respectively. From the above two criteria it is clear that (\ref{eq:BCD and Dbar})
provides us sufficient input for analytic study of the principle squeezing
of vacuum fluctuations, both in single modes and in compound modes.
Similarly, condition for entanglement is in general \cite{Krepelka-1}

\begin{equation}
\left(K_{ij}\right)_{\pm}=(B_{i}\pm|C_{i}|)(B_{j}\pm|C_{j}|)-\left(|D_{ij}|\mp|\bar{D}_{ij}|\right)^{2}<0,\label{eq:entanglement}\end{equation}
and condition for sub-shot noise is \begin{equation}
C_{ij}=B_{i}^{2}+B_{j}^{2}+|C_{i}|^{2}+|C_{j}|^{2}-2|D_{ij}|^{2}-2|\bar{D}_{ij}|^{2}<0;\label{eq:subshot noise}\end{equation}
 further the condition for nonclassical sum- or difference-variance
is \begin{equation}
\left\langle \left(\Delta W_{ij}\right)^{2}\right\rangle _{\pm}=\left\langle \left(\Delta W_{i}\right)^{2}\right\rangle _{N}+\left\langle \left(\Delta W_{j}\right)^{2}\right\rangle _{N}\pm\langle\Delta W_{j}\Delta W_{k}\rangle_{N}<0.\label{eq:varianceplus and minus}\end{equation}
Present work aims to rigorously investigate the presence of different
nonclassicalities in the Raman process in the second order short-time
approximation. To begin with we will discuss intermodal entanglement
in the next section. Before we present our analytic results it is
important to note that for the convenience of understanding the process
we have introduced following two scaled quantities: $gt=\tau$ and
$\frac{|\chi|}{|g|}=p$. The time evolution of various nonclassical
characteristics can now be expressed with respect to dimensionless
quantity $gt=\tau$ and the ratio between the Stokes and anti-Stokes
coupling constants $p$. Further we have used $I_{j}=|\xi_{j}|^{2}$
for the incident stimulating intensities and the phases of the complex
amplitude $\xi_{j}=|\xi_{j}|\exp(\phi_{j})$ are denoted as $\phi_{j}$
that are combined as \[
\phi_{L}-\phi_{V}-\phi_{S}=\theta_{2}\]
 and \[
\phi_{A}-\phi_{L}-\phi_{V}=\theta_{1},\]
where $\theta_{2}$ and $\theta_{1}$ can be visualized as the mismatch
phases in Stokes ($\omega_{S}=\omega_{L}-\omega_{V})$ and in anti-Stokes
($\omega_{A}=\omega_{L}+\omega_{V})$ transitions, respectively. In
the following the coupling constants $g$ and $\chi$ are assumed
to be real.

\section{Phonon mode is coherent\label{sec:Phonon-mode-is-cohrent}}

In the above discussion, all the modes including the phonon mode are
coherent. In such a situation we may investigate the existence of
different kind of nonclassicalities by using (\ref{eq:BCD and Dbar})-(\ref{eq:varianceplus and minus}).
The same is done in the following subsections, where intermode entanglement,
single mode and intermode squeezing, sub-shot noise and variances
are studied in detail.

\subsection{Intermodal entanglement\label{sec:Intermodal-entanglement} }

Substituting (\ref{eq:BCD and Dbar}) in (\ref{eq:entanglement})
we obtain \begin{equation}
\left(K_{LV}\right)_{+}=\left(K_{LV}\right)_{-}=-p^{2}\tau^{2}I_{A}=E_{LV}\label{eq:k-lv}\end{equation}
 and \begin{equation}
\left(K_{SV}\right)_{+}=\left(K_{SV}\right)_{-}=-\tau^{2}I_{L}=E_{SV}.\label{eq:k-sv}\end{equation}
$\left(K_{ij}\right)_{\pm}=0$ for all the other cases. Since we are
using a second order short-time approximated solution we cannot conclude
anything about the separability of those four modes\textbf{\textcolor{blue}{{}
}}for which $\left(K_{ij}\right)_{\pm}=0$. But we can clearly see
that in stimulated Raman process (where $I_{A}\neq0$, $I_{L}\neq0$
, $I_{S}\neq0$, $\, I_{V}\neq0$ ) the vibration-phonon mode is entangled
with the pump-mode and the Stokes mode and it does not depend on $I_{S}$
and $I_{V}$. Consequently if we think of a partially spontaneous
Raman process with $I_{A}\neq0,\, I_{L}\neq0,\, I_{S}=0$, $I_{V}=0,$
then also we will observe both type of photon-phonon entanglement
that we have observed in stimulated Raman process. Interestingly in
the completely spontaneous process (where $I_{A}=0$, $I_{L}\neq0$,
$I_{S}=0$, ~$I_{V}=0$) we can also observe entanglement between
Stokes mode and phonon mode, but in such situation we cannot conclude
about the separability of the pump mode and the phonon mode.

\subsection{Single mode and intermodal squeezing\label{sec:Single-mode-and-squuezing} }

\begin{figure}
\begin{centering}
\includegraphics[scale=0.4]{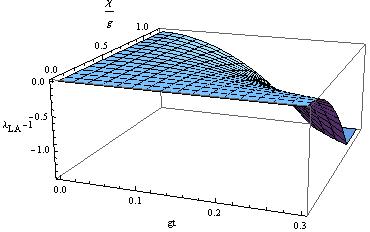}\includegraphics[scale=0.4]{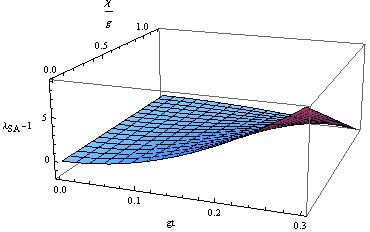}\includegraphics[scale=0.4]{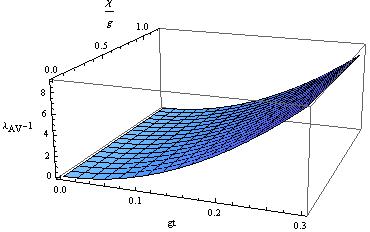}
\par\end{centering}

\begin{centering}
\includegraphics[scale=0.4]{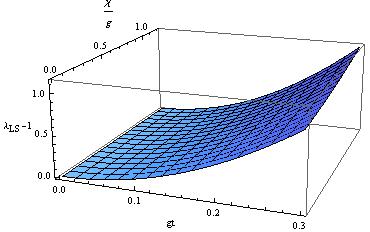}\includegraphics[scale=0.4]{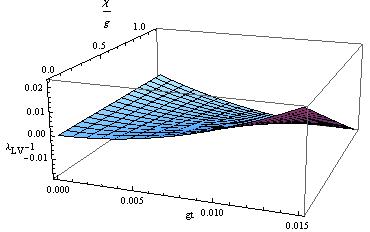}\includegraphics[scale=0.4]{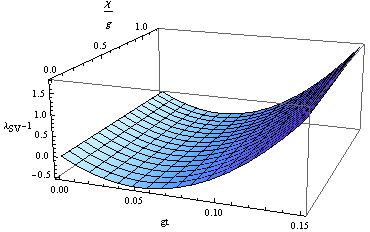}
\par\end{centering}

\caption{\label{fig:Variation-of-lamij}Variation of $\lambda_{ij}-1$ with
respect to $p$ and $gt$ when $|\xi_{L}|=10,\,$ $|\xi_{A}|=1,$
$|\xi_{S}|=9$, $|\xi_{V}|=0.01$ and $\phi_{L}=\phi_{A}$.}
\end{figure}

Substituting (\ref{eq:BCD and Dbar}) in ( \ref{eq:single-mode squuezing})
and (\ref{eq:squeezing compund mode}) we obtain in the interaction
picture reflecting the compensation of exponential function in (\ref{eq:BCD and Dbar})
in a homodyne detection \begin{equation}
\begin{array}{lcl}
\lambda_{L} & = & 1+2p\tau^{2}|\xi_{A}|(p|\xi_{A}|-|\xi_{S}|),\\
\lambda_{LA} & = & 1+p^{2}\tau^{2}|\xi_{A}|\left[|\xi_{A}|-\left(|\xi_{L}|+|\xi_{S}|\right)\right],\\
\lambda_{SA} & = & 1+I_{L}\tau^{2}(1-p),\\
\lambda_{AV} & = & 1+\tau^{2}\left(I_{L}+p^{2}I_{A}-p^{2}|\xi_{A}||\xi_{V}|-p|\xi_{S}||\xi_{A}|\right),\\
\lambda_{LV} & = & 1+2p^{2}\tau^{2}I_{A}+\tau^{2}I_{L}-2\left[p^{2}\tau^{2}I_{A}+\frac{(1+p^{2})^{2}}{4}\tau^{4}I_{L}I_{V}\right]^{\frac{1}{2}}-2p\tau^{2}|\xi_{S}||\xi_{A}|,\\
 & \approx & 1+2p^{2}\tau^{2}I_{A}+\tau^{2}I_{L}-2p\tau I_{A}-2p\tau^{2}|\xi_{S}||\xi_{A}|,\\
\lambda_{LS} & = & 1+\tau^{2}\left(p^{2}I_{A}+I_{L}-|\xi_{L}||\xi_{S}|-p|\xi_{S}||\xi_{A}|+p|\xi_{L}||\xi_{A}|\cos(\phi_{L}-\phi_{A})\right)\\
 & \approx & 1+\tau^{2}I_{L},\\
\lambda_{SV} & = & 1+2\tau^{2}I_{L}+p^{2}\tau^{2}I_{A}-2\left[\tau^{2}I_{L}+\left(\frac{1}{2}\tau^{2}|\xi_{S}||\xi_{V}|+p\tau^{2}|\xi_{A}||\xi_{V}|\right)^{2}\right]^{\frac{1}{2}}+p\tau^{2}|\xi_{S}||\xi_{A}|\\
 & \approx & 1+2\tau^{2}I_{L}+p^{2}\tau^{2}I_{A}-2\tau|\xi_{L}|+p\tau^{2}|\xi_{S}||\xi_{A}|.\end{array}\label{eq:lambda analytic}\end{equation}
From the above equations one can easily obtain the following conditions:
\begin{enumerate}
\item Since $|\xi_{S}|>|\xi_{A}|$ in general the pump mode is always squeezed
if $|g|>|\chi|,$ otherwise it is squeezed if $p|\xi_{A}|<|\xi_{S}|$,
which is expected to be satisfied in most cases. 
\item $\lambda_{LA}<1$ in stimulated Raman process if $|\xi_{A}|<\left(|\xi_{L}|+|\xi_{S}|\right)$,
which is the natural case. So intermodal squeezing between pump and
anti-Stokes mode can be observed in stimulated Raman process. In spontaneous
Raman process $\lambda_{LA}=1$ so squeezing is not observed, but
in partial spontaneous Raman process with $|\xi_{V}|=0,|\xi_{S}|=0,0<|\xi_{A}|<|\xi_{L}|$
squeezing can be observed. 
\item $\lambda_{SA}<1$ iff $p>1,$ i.e. if anti-Stokes coupling is stronger
than the Stokes coupling. If $p>1$ then the intermodal squeezing
in Stokes and anti-Stokes modes is observed for both stimulated and
spontaneous Raman processes.  
\item For a completely spontaneous process $\lambda_{AV}\approx1+\tau^{2}I_{L}$
is always greater than 1. However, also in the stimulated process
the term $\tau^{2}I_{L}$ will be dominant. The same is the\textcolor{red}{{}
}case\textcolor{red}{{} }for $\lambda_{LS}.$
\item For a very short time the linear term in $\lambda_{LV}$ would dominate
and consequently, during that time $\lambda_{LV}\approx1-2p\tau I_{A}$
will be less than unity and consequently squeezing will be observed
in stimulated and partially spontaneous process. 
\item For a very short time the linear term will dominate in $\lambda_{SV}$
and consequently\textcolor{blue}{, }during that time $\lambda_{SV}\approx1-2\tau|\xi_{L}|<1$
would indicate intermodal squeezing in both stimulated and spontaneous
process. 
\end{enumerate}
Variation of $\lambda_{ij}-1$ with respect to $p$ and $\tau$ are
shown in Fig. \ref{fig:Variation-of-lamij}, which clearly depicts
the above observations. %

\subsection{Sub-shot noise\label{sec:Sub-shot-noise} }

When we substitute (\ref{eq:BCD and Dbar}) in (\ref{eq:subshot noise})
we find that \begin{equation}
C_{LV}=-2p^{2}\tau^{2}I_{A}=2(K_{LV})_{\pm}\label{eq:clv}\end{equation}
and \begin{equation}
C_{SV}=-2\tau^{2}I_{A}=2(K_{SV})_{\pm}\label{eq:csv}\end{equation}
and $C_{ij}=0$ in the remaining four cases. Since in this particular
system $C_{LV}$ and $C_{SV}$ are directly proportional to $(K_{LV})_{\pm}$
and $(K_{SV})_{\pm}$, wherever we have seen intermodal entanglement
we, can also observe sub-shot noise in those cases.

\subsection{Variances\label{sec:Variances} }

\begin{figure}
\begin{centering}
\includegraphics[scale=0.5]{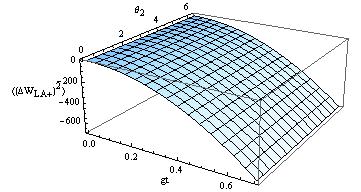}\includegraphics[scale=0.4]{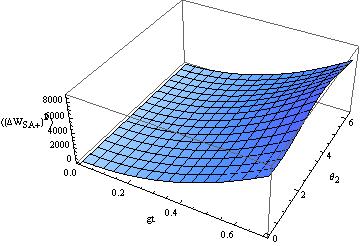}\includegraphics[scale=0.4]{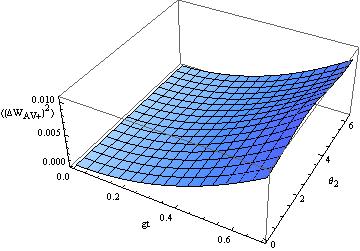}
\par\end{centering}

\begin{centering}
\includegraphics[scale=0.4]{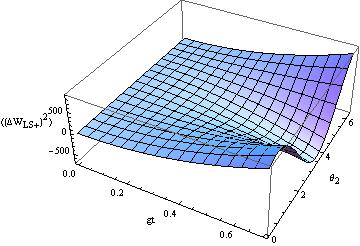}\includegraphics[scale=0.4]{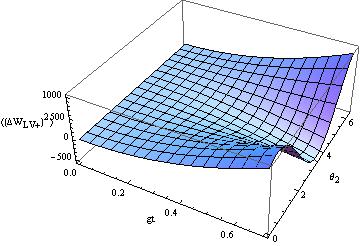}\includegraphics[scale=0.4]{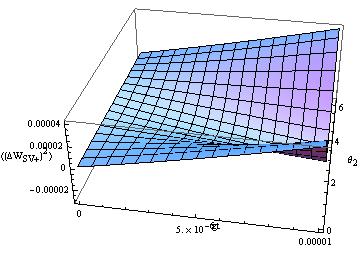}
\par\end{centering}

\caption{\label{fig:wijplus}Variation of intermodal variance $\left\langle (\Delta W)_{ij}^{2}\right\rangle _{+}$
with respect to $gt$ and $\theta_{2}$ here we have chosen $|\xi_{L}|=10,\,$
$|\xi_{A}|=1,$ $|\xi_{S}|=9$, $|\xi_{V}|=0.01,\,$ $\theta_{1}=\frac{\pi}{6}$
and $p=0.9.$}
\end{figure}

\begin{figure}
\begin{centering}
\includegraphics[scale=0.6]{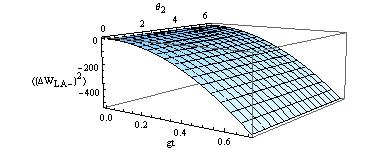}\includegraphics[scale=0.4]{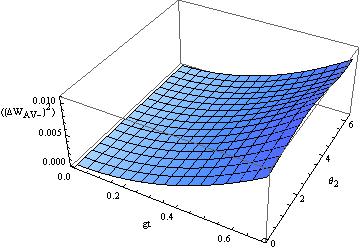}\includegraphics[scale=0.4]{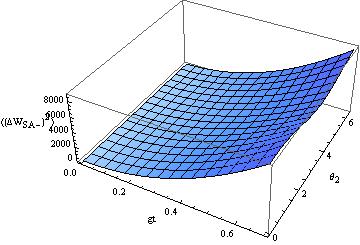}
\par\end{centering}

\begin{centering}
\includegraphics[scale=0.4]{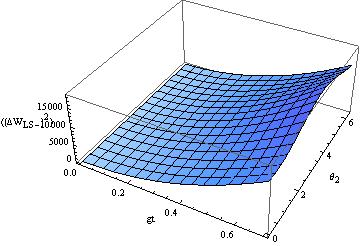}\includegraphics[scale=0.4]{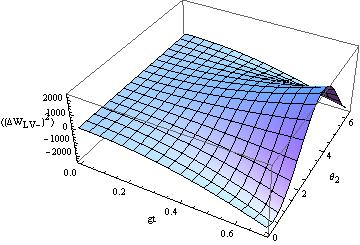}\includegraphics[scale=0.4]{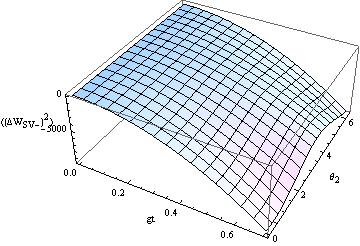}
\par\end{centering}

\caption{\label{fig:wijminus}Variation of variance $\left\langle (\Delta W)_{ij}^{2}\right\rangle _{-}$
with respect to $gt$ and $\theta_{2}$ here we have chosen $|\xi_{L}|=10,\,$
$|\xi_{A}|=1,$ $|\xi_{S}|=9$, $|\xi_{V}|=0.01,\,$ $\theta_{1}=\frac{\pi}{6},$
and $p=0.9.$}
\end{figure}

Using (\ref{eq:BCD and Dbar})-(\ref{eq:cross variance}) and (\ref{eq:varianceplus and minus})
we can obtain the analytic expressions for intermodal variances in
the following forms

\begin{equation}
\begin{array}{lcl}
\left\langle \left(\Delta W_{AV}\right)^{2}\right\rangle _{+} & = & 2\tau^{2}I_{V}\left(I_{L}-p|\xi_{S}||\xi_{A}|\cos(\theta_{1}+\theta_{2})\right),\\
\left\langle \left(\Delta W_{SA}\right)^{2}\right\rangle _{+} & = & 2\tau^{2}I_{L}\left(I_{S}-p|\xi_{S}||\xi_{A}|\cos(\theta_{2}-\theta_{1})\right),\\
\left\langle \left(\Delta W_{LA}\right)^{2}\right\rangle _{+} & = & -2p\tau^{2}I_{L}|\xi_{S}||\xi_{A}|\cos(\theta_{2}-\theta_{1}),\\
\left\langle \left(\Delta W_{LS}\right)^{2}\right\rangle _{+} & = & 2p\tau^{2}I_{L}|\xi_{A}|\left(p|\xi_{A}|+|\xi_{S}|\cos(\theta_{2}-\theta_{1})\right),\\
\left\langle \left(\Delta W_{LV}\right)^{2}\right\rangle _{+} & = & 2p^{2}\tau^{2}\left(3I_{L}I_{A}+3I_{A}I_{V}+I_{A}-I_{L}I_{V}\right)\\
 & + & 2p\tau^{2}|\xi_{S}||\xi_{A}|\left(I_{L}\cos(\theta_{2}-\theta_{1})+I_{V}\cos(\theta_{1}+\theta_{2})\right)+4p\tau|\xi_{A}||\xi_{L}||\xi_{V}|\cos(\theta_{1}),\\
\left\langle \left(\Delta W_{SV}\right)^{2}\right\rangle _{+} & = & 2\tau^{2}\left(3I_{L}I_{S}+3I_{L}I_{V}+p^{2}I_{A}I_{V}+I_{L}-I_{S}I_{V}\right)\\
 & + & 2p\tau^{2}|\xi_{S}||\xi_{A}|\left(2I_{L}\cos(\theta_{2}-\theta_{1})-3I_{V}\cos(\theta_{1}+\theta_{2})\right)+4\tau|\xi_{L}||\xi_{S}||\xi_{V}|\cos(\theta_{2}),\end{array}\label{eq:wplus}\end{equation}
and 

\begin{equation}
\begin{array}{lcl}
\left\langle \left(\Delta W_{AV}\right)^{2}\right\rangle _{-} & = & 2\tau^{2}I_{V}\left(I_{L}+2p^{2}I_{A}-p|\xi_{S}||\xi_{A}|\cos(\theta_{1}+\theta_{2})\right),\\
\left\langle \left(\Delta W_{SA}\right)^{2}\right\rangle _{-} & = & 2\tau^{2}I_{L}\left(I_{S}+p|\xi_{S}||\xi_{A}|\cos(\theta_{2}-\theta_{1})\right),\\
\left\langle \left(\Delta W_{LA}\right)^{2}\right\rangle _{-} & = & 2p\tau^{2}I_{L}\left(2pI_{A}-|\xi_{S}||\xi_{A}|\cos(\theta_{2}-\theta_{1})\right),\\
\left\langle \left(\Delta W_{LS}\right)^{2}\right\rangle _{-} & = & 2\tau^{2}I_{L}\left(p^{2}I_{A}+2I_{S}-3p|\xi_{S}||\xi_{A}|\cos(\theta_{2}-\theta_{1})\right),\\
\left\langle \left(\Delta W_{LV}\right)^{2}\right\rangle _{-} & = & -2p^{2}\tau^{2}\left(I_{L}I_{A}+I_{A}I_{V}+I_{A}-I_{L}I_{V}\right)+4\tau^{2}I_{L}I_{V}\\
 & - & 6p\tau^{2}|\xi_{S}||\xi_{A}|\left(I_{L}\cos(\theta_{2}-\theta_{1})+I_{V}\cos(\theta_{1}+\theta_{2})\right)-4p\tau|\xi_{A}||\xi_{L}||\xi_{V}|\cos(\theta_{1}),\\
\left\langle \left(\Delta W_{SV}\right)^{2}\right\rangle _{-} & = & -2\tau^{2}\left(I_{L}I_{S}+I_{L}I_{V}-p^{2}I_{A}I_{V}+I_{L}-I_{S}I_{V}\right)\\
 & - & 2p\tau^{2}|\xi_{S}||\xi_{A}|\left(2I_{L}\cos(\theta_{2}-\theta_{1})-I_{V}\cos(\theta_{1}+\theta_{2})\right)-4\tau|\xi_{L}||\xi_{S}||\xi_{V}|\cos(\theta_{2}).\end{array}\label{eq:wminus}\end{equation}
Negativity of intermodal variances $\left\langle (\Delta W)_{ij}^{2}\right\rangle _{\pm}$
implies nonclassicality. Analytic expressions for intermodal variances
$\left\langle (\Delta W)_{ij}^{2}\right\rangle _{+}$ and $\left\langle (\Delta W)_{ij}^{2}\right\rangle _{-}$
for all the possible combinations of modes in the stimulated Raman
process are provided in (\ref{eq:wplus}) and (\ref{eq:wminus}),
respectively. It is difficult to conclude directly about the presence
of nonclassicality from these general analytic expressions of $\left\langle (\Delta W)_{ij}^{2}\right\rangle _{\pm}.$
Thus to visualize the existence of nonclassicality we have plotted
the analytic expressions provided in (\ref{eq:wplus}) and (\ref{eq:wminus}).
The plots are given in Fig. \ref{fig:wijplus} and Fig. \ref{fig:wijminus}
and it is easy to see that both $\left\langle (\Delta W)_{ij}^{2}\right\rangle _{+}$
and $\left\langle (\Delta W)_{ij}^{2}\right\rangle _{-}$ depicts
nonclassical behavior for a) pump and phonon mode and b) pump and
anti-Stokes mode, c) Stokes and phonon mode. However for pump and
Stokes mode only $\left\langle (\Delta W)_{LS}^{2}\right\rangle _{+}$
shows the existence of nonclassicality. 

For the chosen values of $|\xi_{i}|$ and $p$ we have seen that $\left\langle (\Delta W)_{SA}^{2}\right\rangle _{\pm}=2\tau^{2}I_{L}|\xi_{S}|\left(|\xi_{S}|\pm p|\xi_{A}|\cos(\theta_{2}-\theta_{1})\right)>0$
but a negative value is possible if $p|\xi_{A}|>|\xi_{S}|$. Thus
a very strong anti-Stokes coupling (compared to Stokes coupling) may
yield nonclassical variance for Stokes and anti-Stokes mode. This
is consistent with the appearance of intermodal squeezing where $\lambda_{AS}-1=I_{L}\tau^{2}(1-p)$
is negative only when $p>1,$ that is when anti-Stokes coupling is
stronger than Stokes coupling. 

Now from (\ref{eq:wminus}) we can easily observe that for a completely
spontaneous Raman process $\left\langle (\Delta W)_{SV}^{2}\right\rangle _{-}=-2\tau^{2}I_{L}$
is always negative which indicates intermodal nonclassical behavior
between phonon mode and Stokes mode. We have already shown that these
two modes show intermodal entanglement, sub-shot noise behavior and
squeezing of vacuum fluctuations in the spontaneous Raman process.
Thus as far as the nonclassicalities in spontaneous Raman process
are concerned these two modes play the most important role. 

\begin{figure}
\begin{centering}
\includegraphics[scale=0.4]{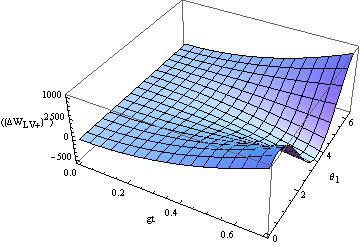}\includegraphics[scale=0.4]{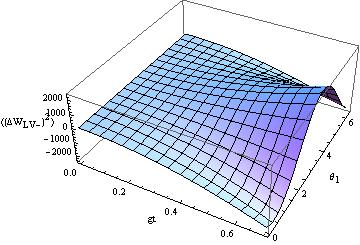}
\par\end{centering}

\caption{\label{fig:Variation-of-wlv-shorttime}Variation of variance $\left\langle (\Delta W)_{LV}^{2}\right\rangle _{\pm}$
with respect to $\theta_{1}$ when $\theta_{2}=\frac{\pi}{6}$ all
other parameters are same as in previous figures.}
\end{figure}

From (\ref{eq:wplus}) and (\ref{eq:wminus}) we can see that for
very small values of rescaled time $\tau$ the term linear in $\tau$
is expected to dominate in $\left\langle (\Delta W)_{LV}^{2}\right\rangle _{\pm}$
and in $\left\langle (\Delta W)_{SV}^{2}\right\rangle _{\pm}$; $\left\langle (\Delta W)_{LV}^{2}\right\rangle _{\pm}$
varies with $\theta_{1}$, which is exhibited in Fig. \ref{fig:Variation-of-wlv-shorttime}.
Further the linear term in $\left\langle (\Delta W)_{SV}^{2}\right\rangle _{+}$
is very weak and the nonclassical behavior can be seen only for a
very small values of $\tau$. This is why in Fig. \ref{fig:wijplus}
we have plotted $\left\langle (\Delta W)_{SV}^{2}\right\rangle _{+}$
for a very short time only. 

\section{Phonon mode is chaotic\label{sec:Phonon-mode-is-chaotic} }

In this case we perform the average over the initial phonon amplitude
in (\ref{eq:characteristic function}) with a Gaussian distribution
in the Gaussian approximation. Assuming that the phonon mode is chaotic
with average phonon number $\langle n_{V}\rangle,$ then the coefficients
in the interaction picture described in (\ref{eq:BCD and Dbar}) get
modified as \begin{equation}
\begin{array}{lcl}
B_{L} & = & |\chi|^{2}t^{2}|\xi_{A}|^{2}\left(\langle n_{V}\rangle+1\right)+|g|^{2}t^{2}|\xi_{S}|^{2}\langle n_{V}\rangle,\\
B_{S} & = & |g|^{2}t^{2}|\xi_{L}|^{2}\left(\langle n_{V}\rangle+1\right),\\
B_{A} & = & |\chi|^{2}t^{2}|\xi_{L}|^{2}\langle n_{V}\rangle,\\
B_{V} & \approx & \langle n_{V}\rangle,\\
C_{L} & = & -g^{*}\chi t^{2}\xi_{A}\xi_{S}\left(2\langle n_{V}\rangle+1\right),\\
C_{S} & = & 0,\\
C_{A} & = & 0,\\
C_{V} & = & -g\chi t^{2}\xi_{S}^{*}\xi_{A},\\
D_{LS} & = & -\frac{1}{2}|g|^{2}t^{2}\xi_{L}\xi_{S}\left(2\langle n_{V}\rangle+1\right),\\
D_{LA} & = & -\frac{1}{2}|\chi|^{2}t^{2}\xi_{L}\xi_{A}\left(2\langle n_{V}\rangle+1\right),\\
D_{SA} & = & -\frac{1}{2}g\chi^{*}t^{2}\xi_{L}^{2}\left(2\langle n_{V}\rangle+1\right),\\
D_{SV} & = & igt\xi_{L}\left(\langle n_{V}\rangle+1\right),\\
D_{LV} & = & i\chi t\xi_{A}\left(\langle n_{V}\rangle+1\right),\\
D_{AV} & = & 0,\\
\bar{D}_{LS} & = & -g\chi^{*}t^{2}\xi_{A}^{*}\xi_{L}\left(\langle n_{V}\rangle+1\right),\\
\bar{D}_{LA} & = & -g\chi^{*}t^{2}\xi_{S}^{*}\xi_{L}\langle n_{V}\rangle,\\
\bar{D}_{SA} & = & 0,\\
\bar{D}_{SV} & = & 0,\\
\bar{D}_{LV} & = & igt\xi_{S}^{*}\langle n_{V}\rangle,\\
\bar{D}_{AV} & = & i\chi t\xi_{L}^{*}\langle n_{V}\rangle.\end{array}\label{eq:chaotic-bcdbar}\end{equation}
Now using equations (\ref{eq:variance}), (\ref{eq:cross variance})
and (\ref{eq:chaotic-bcdbar}) we can obtain for single-mode variances
\begin{equation}
\begin{array}{lcl}
\langle\left(\Delta W_{L}\right)^{2}\rangle_{N} & = & 2p^{2}\tau^{2}I_{A}I_{L}\left(\langle n_{V}\rangle+1\right)+2\tau^{2}I_{S}I_{L}\langle n_{V}\rangle-2p\tau^{2}|\xi_{A}||\xi_{S}|I_{L}\left(2\langle n_{V}\rangle+1\right)\cos(\theta_{2}-\theta_{1}),\\
\langle\left(\Delta W_{S}\right)^{2}\rangle_{N} & = & 2\tau^{2}I_{L}I_{S}\left(\langle n_{V}\rangle+1\right),\\
\langle\left(\Delta W_{A}\right)^{2}\rangle_{N} & = & 2p^{2}\tau^{2}I_{L}I_{A}\langle n_{V}\rangle,\\
\langle\left(\Delta W_{V}\right)^{2}\rangle_{N} & \approx & \langle n_{V}\rangle^{2},\end{array}\label{eq:chaotic w-square}\end{equation}
and for correlation fluctuations \begin{equation}
\begin{array}{lcl}
\left(\Delta W_{L}\Delta W_{A}\right)_{N} & = & -\tau^{2}p^{2}I_{L}I_{A}\left(2\langle n_{V}\rangle+1\right)+2p\tau^{2}I_{L}|\xi_{S}||\xi_{A}|\langle n_{V}\rangle\cos(\theta_{2}-\theta_{1}),\\
\left(\Delta W_{L}\Delta W_{S}\right)_{N} & = & -\tau^{2}I_{L}I_{S}\left(2\langle n_{V}\rangle+1\right)+2p\tau^{2}I_{L}|\xi_{S}||\xi_{A}|\left(\langle n_{V}\rangle+1\right)\cos(\theta_{2}-\theta_{1}),\\
\left(\Delta W_{S}\Delta W_{A}\right)_{N} & = & -\tau^{2}pI_{L}|\xi_{A}||\xi_{S}|\cos(\theta_{2}-\theta_{1})\left(2\langle n_{V}\rangle+1\right).\end{array}\label{eq:chaotic cross correln}\end{equation}
Analytic expressions of the other cross-correlations are not of interest
as all variances that involve phonon mode will always be positive
because of the dominance of $\langle\left(\Delta W_{V}\right)^{2}\rangle\approx\langle n_{V}\rangle^{2}$
term. Now substituting equations (\ref{eq:chaotic-bcdbar}), (\ref{eq:chaotic w-square})
and (\ref{eq:chaotic cross correln}) in the criteria of nonclassicalities
introduced in (\ref{eq:single-mode squuezing})-(\ref{eq:varianceplus and minus})
we can investigate the nonclassical character of stimulated and spontaneous
Raman process when the phonon mode is chaotic and then compare the
results with the similar results obtained in the coherent case. This
is done in the following subsections.

\subsection{Intermodal entanglement}

Substituting (\ref{eq:chaotic-bcdbar}) in (\ref{eq:entanglement})
we obtain \begin{equation}
\begin{array}{lcl}
\left(K_{SV}\right)_{\pm} & = & -\tau^{2}I_{L}\left(\langle n_{V}\rangle+1\right),\\
\left(K_{LV}\right)_{\pm} & = & -p^{2}\tau^{2}I_{A}\left(\langle n_{V}\rangle+1\right)\mp2p\tau^{2}|\xi_{A}||\xi_{S}|\langle n_{V}\rangle\left(3\langle n_{V}\rangle+2\right).\end{array}\label{eq:chaotic-kplusminus}\end{equation}
All other $\left(K_{ij}\right)_{\pm}=0$. 

We can conclude:
\begin{enumerate}
\item From (\ref{eq:chaotic-kplusminus}) it is clear that the phonon mode
is always entangled with Stokes mode. The same characteristic was
also observed in coherent case but if we consider $\left(K_{SV}\right)_{\pm}$
as a measure of amount of entanglement, then the amount of entanglement
in chaotic case is increased by a factor of $\left(1+\langle n_{V}\rangle\right)$
and it is more announced. 
\item Similarly the phonon mode can be entangled with the pump mode. It
is straightforward to see that $\left(K_{LV}\right)_{+}=-p^{2}\tau^{2}I_{A}\left(\langle n_{V}\rangle+1\right)-2p\tau^{2}|\xi_{A}||\xi_{S}|\langle n_{V}\rangle\left(\left(3\langle n_{V}\rangle+2\right)\right)$
exhibits intermodal entanglement. 
\item But interestingly $\left(K_{LV}\right)_{-}$ does not show signature
of intermodal entanglement. 
\item Stokes mode and phonon mode are entangled for completely spontaneous
Raman process also but the present calculation is non-conclusive about
entanglement of pump and phonon mode. 
\end{enumerate}

\subsection{Single mode and intermodal squeezing }

By substituting (\ref{eq:chaotic-bcdbar}) in (\ref{eq:single-mode squuezing})
and (\ref{eq:squeezing compund mode}) we obtain \begin{equation}
\begin{array}{lcl}
\lambda_{L} & = & 1+2\tau^{2}\left[p^{2}I_{A}\left(\langle n_{V}\rangle+1\right)+I_{S}\langle n_{V}\rangle-p|\xi_{A}||\xi_{S}|\left(2\langle n_{V}\rangle+1\right)\right],\\
\lambda_{LA} & \approx & 1+p^{2}\tau^{2}|\xi_{L}|\left(|\xi_{L}|\langle n_{V}\rangle-|\xi_{A}|\left(2\langle n_{V}\rangle+1\right)\right),\\
\lambda_{LS} & \approx & 1+\tau^{2}|\xi_{L}|\left(|\xi_{L}|\left(\langle n_{V}\rangle+1\right)-|\xi_{S}|\left(2\langle n_{V}\rangle+1\right)\right),\\
\lambda_{SA} & = & 1+\tau^{2}I_{L}(1-p)\left(\langle n_{V}\rangle+1-p\langle n_{V}\rangle\right).\end{array}\label{eq:pustak1}\end{equation}
 We see that:
\begin{enumerate}
\item Squeezing in the pump laser mode can be observed by approximating
$\frac{|\xi_{A}|}{|\xi_{S}|}>\frac{\langle n_{V}\rangle}{\left(2\langle n_{V}\rangle+1\right)}$. 
\item Intermodal squeezing is not possible when one of the mode is phonon
mode as in that case $\lambda_{iV}\approx1+\langle n_{V}\rangle>1.$
\item Intermodal squeezing will not be usually observed between pump mode
and anti-Stokes mode as $\lambda_{LA}<1$ implies $|\xi_{L}|<2|\xi_{A}|+\frac{|\xi_{A}|}{\langle n_{V}\rangle}\approx2|\xi_{A}|$.
But technically it is allowed and intermodal squeezing between pump
mode and anti-Stokes mode can in principle be seen for stimulated
Raman process as well as for partially spontaneous Raman process ($I_{A}\neq0$,
$I_{L}\neq0$, $I_{S}=0$, ~ $I_{V}=0$). 
\item Intermodal squeezing between pump mode and Stokes mode is possible
if $|\xi_{L}|\left(\langle n_{V}\rangle+1\right)<|\xi_{S}|\left(2\langle n_{V}\rangle+1\right),$
i.e. if $\frac{|\xi_{L}|}{|\xi_{S}|}<\frac{\left(2\langle n_{V}\rangle+1\right)}{\left(\langle n_{V}\rangle+1\right)}$.
For $\langle n_{V}\rangle\gg1$ this condition implies that $|\xi_{L}|<2|\xi_{S}|$
and for $\langle n_{V}\rangle=0$ it implies $|\xi_{L}|<|\xi_{S}|.$
\item To have $\lambda_{SA}<1,$ we need $p=\frac{|\chi|}{|g|}>1$ and $\left(\langle n_{V}\rangle+1\right)>p\langle n_{V}\rangle,$
i.e. $\frac{|\chi|}{|g|}<1+\frac{1}{\langle n_{V}\rangle}.$ For coherent
scattering $\langle n_{V}\rangle\rightarrow0$ and we have the previous
condition $\frac{|\chi|}{|g|}>1.$
\item Similarly when $p<1$ then $\lambda_{SA}<1$ implies $\left(p\langle n_{V}\rangle-\left(\langle n_{V}\rangle+1\right)\right)>0\Rightarrow p>1+\frac{1}{\langle n_{V}\rangle}\Rightarrow p>1$.
Thus the condition of negativity is not satisfied and we are non-conclusive
about the entanglement between Stokes mode and anti-Stokes mode if
$p<1.$ 
\end{enumerate}

\subsection{Sub-shot noise }

By substituting (\ref{eq:chaotic-bcdbar}) in (\ref{eq:subshot noise})
we obtain \begin{equation}
\begin{array}{lcl}
C_{AV} & = & \langle n_{V}\rangle^{2}\left(1-2p^{2}\tau^{2}I_{L}\right),\\
C_{LV} & = & \langle n_{V}\rangle^{2}-2p^{2}\tau^{2}I_{A}\left(\langle n_{V}\rangle+1\right)^{2}-2\tau^{2}I_{S}\langle n_{V}\rangle^{2},\\
C_{SV} & = & \langle n_{V}\rangle^{2}-2\tau^{2}I_{L}\left(\langle n_{V}\rangle+1\right)^{2},\end{array}\label{eq:pustak2}\end{equation}
and all other $C_{ij}=0.$ For stimulated Raman process, sub-shot
noise is observed in the above three cases. In coherent case subshot
noise behavior was not observed for anti-Stokes and phonon mode. Further,
negativity of $C_{AV}$ and $C_{SV}$ will be observed for spontaneous
Raman process too. But in the spontaneous Raman process sub-shot noise
behavior will not be observed for pump and phonon modes. However,
we can observe it for partially spontaneous process ($\langle n_{V}\rangle\neq0,\, I_{L}\neq0,\, I_{A}=0$
and $I_{S}=0$.)

\subsection{Variances }

By substituting (\ref{eq:chaotic w-square}) and (\ref{eq:chaotic cross correln})
in (\ref{eq:varianceplus and minus}) we obtain \begin{equation}
\begin{array}{lcl}
\langle\left(\Delta W\right)_{SA}^{2}\rangle_{\pm} & = & 2p^{2}\tau^{2}I_{L}I_{A}\langle n_{V}\rangle+2\tau^{2}I_{L}I_{S}\left(\langle n_{V}\rangle+1\right)\\
 & \mp & 2\tau^{2}pI_{L}|\xi_{A}||\xi_{S}|\cos(\theta_{2}-\theta_{1})\left(2\langle n_{V}\rangle+1\right),\\
\langle\left(\Delta W\right)_{LS}^{2}\rangle_{\pm} & = & 2p^{2}\tau^{2}I_{A}I_{L}\left(\langle n_{V}\rangle+1\right)+2\tau^{2}I_{L}I_{S}\left(2\langle n_{V}\rangle+1\right)(1\mp1)\\
 & - & 2p\tau^{2}I_{L}|\xi_{S}||\xi_{A}|\cos(\theta_{2}-\theta_{1})\left(\left(2\langle n_{V}\rangle+1\right)\mp2\left(\langle n_{V}\rangle+1\right)\right),\\
\langle\left(\Delta W\right)_{LA}^{2}\rangle_{\pm} & = & 2p^{2}\tau^{2}I_{A}I_{L}\left(\left(2\langle n_{V}\rangle+1\right)\mp\left(\langle n_{V}\rangle+1\right)\right)+2\tau^{2}I_{S}I_{L}\langle n_{V}\rangle\\
 & - & 2p\tau^{2}|\xi_{A}||\xi_{S}|I_{L}\cos(\theta_{2}-\theta_{1})\left(\left(2\langle n_{V}\rangle+1\right)\mp2\langle n_{V}\rangle\right).\end{array}\label{eq:chaotic variance}\end{equation}
From (\ref{eq:chaotic variance}) we observe following: 
\begin{enumerate}
\item $\langle\left(\Delta W\right)_{LS}^{2}\rangle_{+}=2p\tau^{2}|\xi_{A}|I_{L}\left(p|\xi_{A}|\left(\langle n_{V}\rangle+1\right)+|\xi_{S}|\cos(\theta_{2}-\theta_{1})\right)$.
Thus negative variance can be seen for $p|\xi_{A}|\left(\langle n_{V}\rangle+1\right)<|\xi_{S}|$.
Since $|\xi_{A}|<|\xi_{S}|$ this nonclassical feature between pump
mode and Stokes mode will be observed for small values of mean phonon
number $\langle n_{V}\rangle.$
\item In the analytic expression of $\langle\left(\Delta W\right)_{AS}^{2}\rangle_{\pm}$
if we assume $I_{A}\ll I_{S}$ then we obtain \[
\langle\left(\Delta W\right)_{AS}^{2}\rangle_{\pm}=2\tau^{2}I_{L}|\xi_{S}|\left(|\xi_{S}|\left(\langle n_{V}\rangle+1\right)\mp p|\xi_{A}|\left(2\langle n_{V}\rangle+1\right)\cos(\theta_{2}-\theta_{1})\right)\]
which would show nonclassicality if $p\frac{|\xi_{A}|}{|\xi_{S}|}>\frac{\langle n_{V}\rangle+1}{2\langle n_{V}\rangle+1}$.
This implies $p|\xi_{A}|>\frac{1}{2}|\xi_{S}|$ which is inconsistent
with the assumption $I_{A}\ll I_{S}$. Thus if $I_{A}\ll I_{S}$ then
we do not observe nonclassical variance in anti-Stokes and Stokes
modes. 
\item If we assume $I_{A}\gg I_{S}$  and consider the complete analytic
expression of $\langle\left(\Delta W_{AS}\right)^{2}\rangle_{\pm},$
then the condition  $p\frac{|\xi_{A}|}{|\xi_{S}|}>\frac{\langle n_{V}\rangle+1}{2\langle n_{V}\rangle+1}$
will serve as necessary but not sufficient condition of nonclassicality.
Now if we assume that for some choice of $p,I_{A},I_{S},\langle n_{V}\rangle$
we observe nonclassical intermodal variance for Stokes and anti-Stokes
modes, then we can show that for that situation $\langle\left(\Delta W_{LS}\right)^{2}\rangle_{+}$
will not show nonclassicality. The proof is simple. First we assume
that both $\langle\left(\Delta W_{LS}\right)^{2}\rangle_{+}$ and
$\langle\left(\Delta W_{AS}\right)^{2}\rangle_{\pm}$ are negative.
Therefore, $\frac{1}{\langle n_{V}\rangle+1}>p\frac{|\xi_{A}|}{|\xi_{S}|}>\frac{\langle n_{V}\rangle+1}{2\langle n_{V}\rangle+1}$,
which implies $\left(2\langle n_{V}\rangle+1\right)>\left(\langle n_{V}\rangle+1\right)^{2}$
or $\langle n_{V}\rangle^{2}<0$. Thus by reductio ad absurdum we
have shown that intermodal nonclassical variance cannot be seen simultaneously
in a) Stokes and anti-Stokes mode and b) Stokes and pump mode. 
\item For the compound mode $(LA)$ one could observe sub-shot noise provided
that $|\chi||\xi_{A}|>\frac{|g||\xi_{S}|}{\sqrt{2}}.$
\end{enumerate}

\section{Joint photon-phonon number and wave distribution\label{sec:Joint-photon-phonon-number}}

\begin{figure}
\begin{centering}
\includegraphics[scale=0.5]{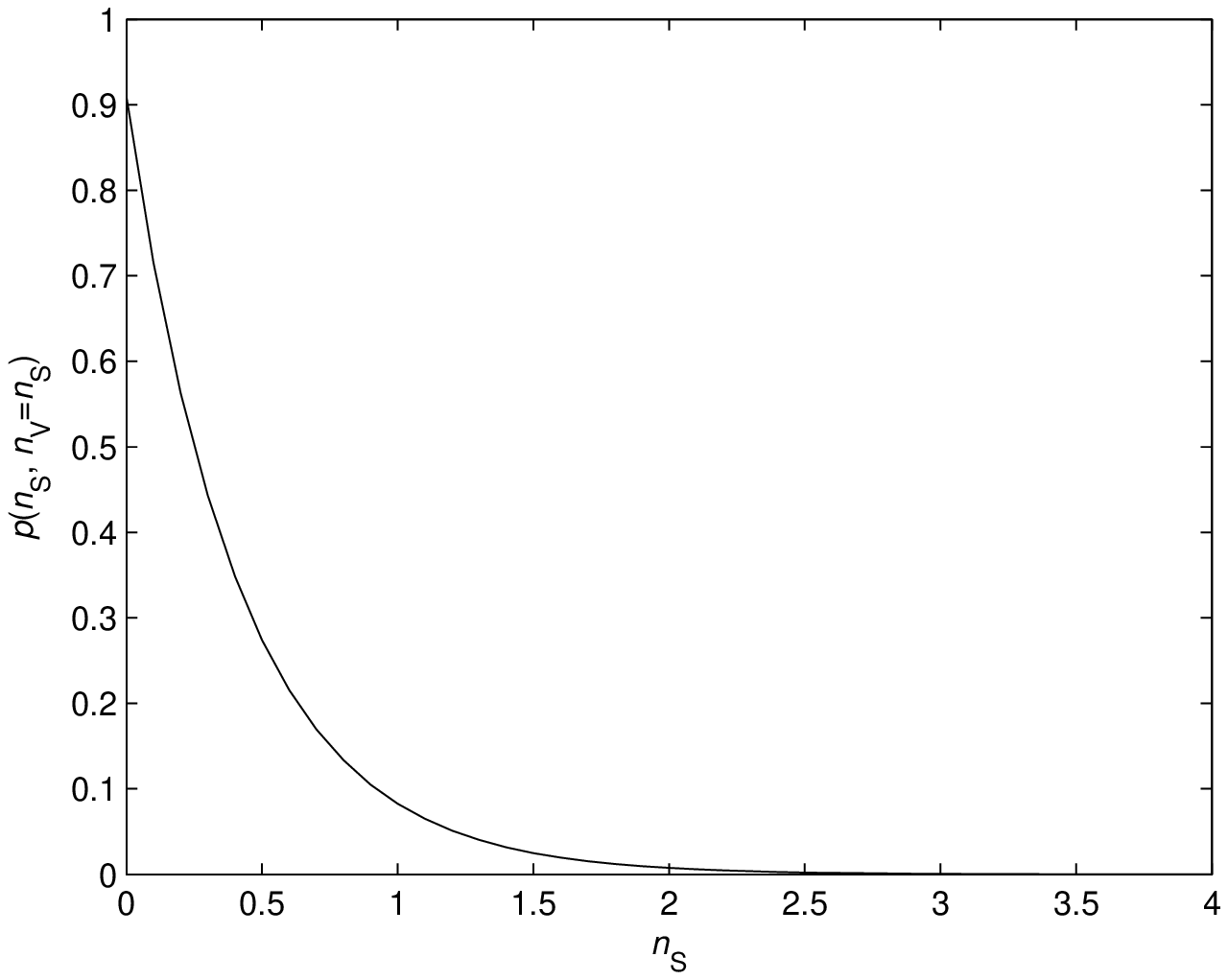}\includegraphics[scale=0.5]{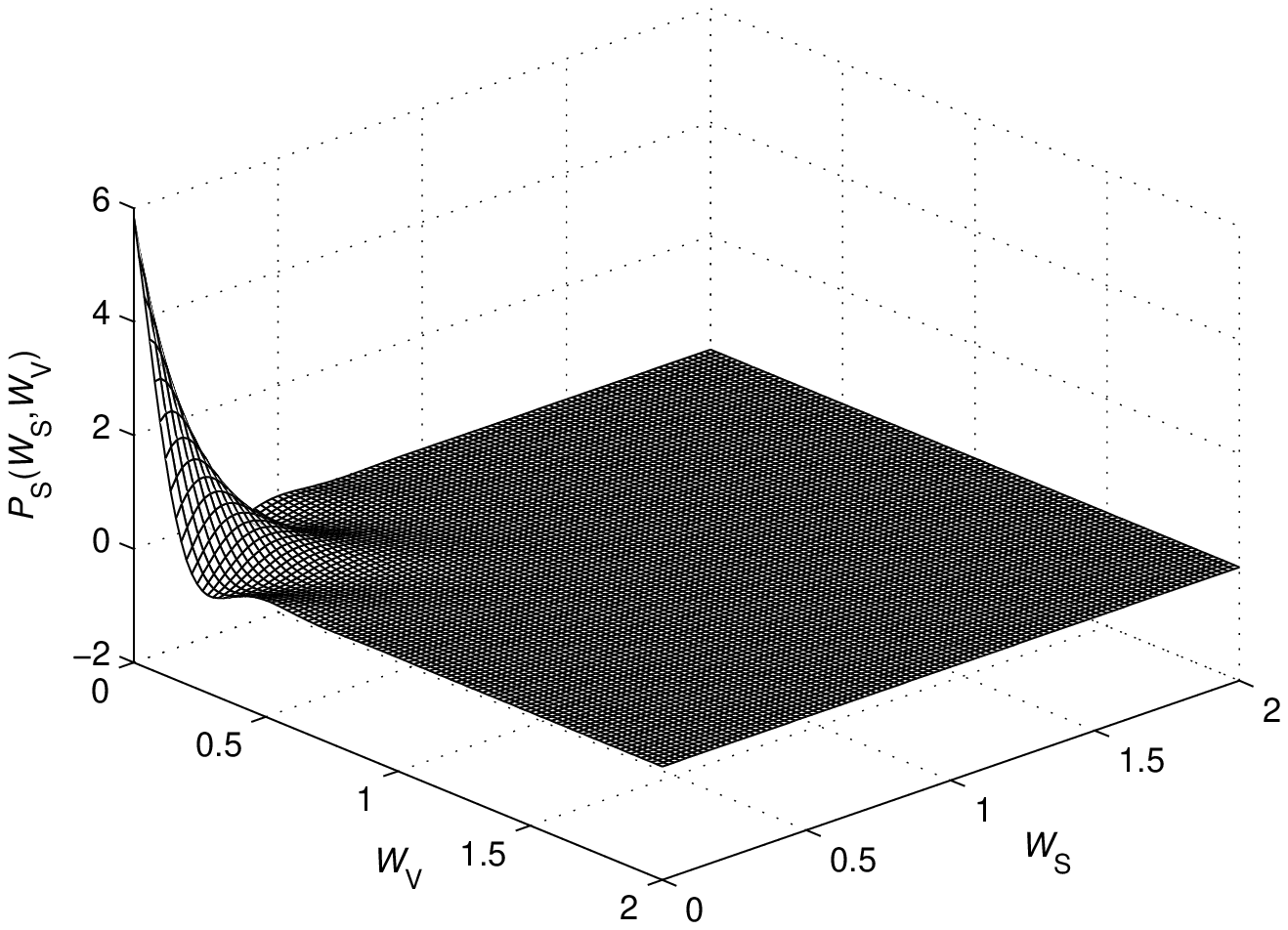}
\par\end{centering}

\caption{\label{fig:pnsnv} a) Joint photon-phonon number distribution for
Stokes and vibration modes (left) and b) Quasi-distribution of integrated-intensities
for the same modes (right). Here $B_{S}=0.1,\, B_{L}=0.01,\, B_{V}=0.11$
and $s=0.7.$ }
\end{figure}

We can illustrate the above results for nonclassical behavior of modes
in Raman scattering by joint photon-phonon number and integrated-intensity
distributions along the lines given in \cite{Krepelka-1} (and references
therein) in Gaussian approximation. For simplicity we consider scattering
by phonon vacuum (in optical region and for room temperature $\langle n_{V}\rangle\approx0$)
for compound modes $(SV$ ) and $(LV)$, which exhibit quantum entanglement
up to the second order in $t$. From (\ref{eq:BCD and Dbar}) we see
that $K_{SV}=-B_{S}=-B_{V}=-|g|^{2}t^{2}I_{L}$ provided that we consider
spontaneous scattering $(I_{A}=I_{S}=0;$ in this case $C_{S}=C_{V}=\bar{D}_{SV}=0).$
In principle we can also consider partially stimulated scattering
with $I_{S}\neq0,$ when using shifted distributions in $W_{S}$ along
$I_{S}$ \cite{Krepelka-2} to adopt spontaneous process. Thus $K_{SV}+B_{S}=K_{SV}+B_{V}=0$
and from the formulae given in \cite{Krepelka-1} we obtain the joint
photon-phonon number distribution \begin{equation}
p(n_{S},n_{V})=\frac{(B_{S})^{n_{S}}}{(1+B_{S})^{1+n_{S}}}\delta_{n_{S},n_{V}},\label{eq:pustak3}\end{equation}
i.e. it is diagonal expressing a pairwise structure of photon-phonon
process in this case. It is shown in Fig. \ref{fig:pnsnv}a. The corresponding
$s$-order quasidistribution of integrated-intensities is \cite{Krepelka-1}
\begin{equation}
P_{s}\left(W_{S},W_{V}\right)=\frac{1}{\pi B_{Ss}}{\rm e}^{-\frac{W_{S}+W_{V}}{2B_{Ss}}}\frac{\sin\left(\frac{W_{S}-W_{V}}{\sqrt{-K_{VS,s}}}\right)}{W_{S}-W_{V}},\label{eq:pustak4}\end{equation}
where $B_{Ss}=B_{S}+\frac{1-s}{2}$ and $K_{SV,s}=K_{SV}+(1-s)B_{S}+\frac{(1-s)^{2}}{4};$
$s$ is ordering parameter. For the threshold value of the ordering
parameter we have $s_{th}=1+B_{S}+B_{V}-\sqrt{(B_{S}+B_{V})^{2}-4K_{SV}}\approx1+2B_{S}-2\sqrt{B_{S}}.$
Choosing $|g|t=0.1,\, I_{L}=10,$ we have $s_{th}=0.57.$ So we calculate
the quasi-distribution for $s=0.7;$ in this case $B_{S}=0.1,$ $B_{Ss}=0.25,$
$K_{VS,s}=-0.048.$ This quasi-distribution is shown in Fig. \ref{fig:pnsnv}b.
It takes on negative values exhibiting nonclassical oscillations and
behavior.

\begin{figure}
\begin{centering}
\includegraphics[scale=0.5]{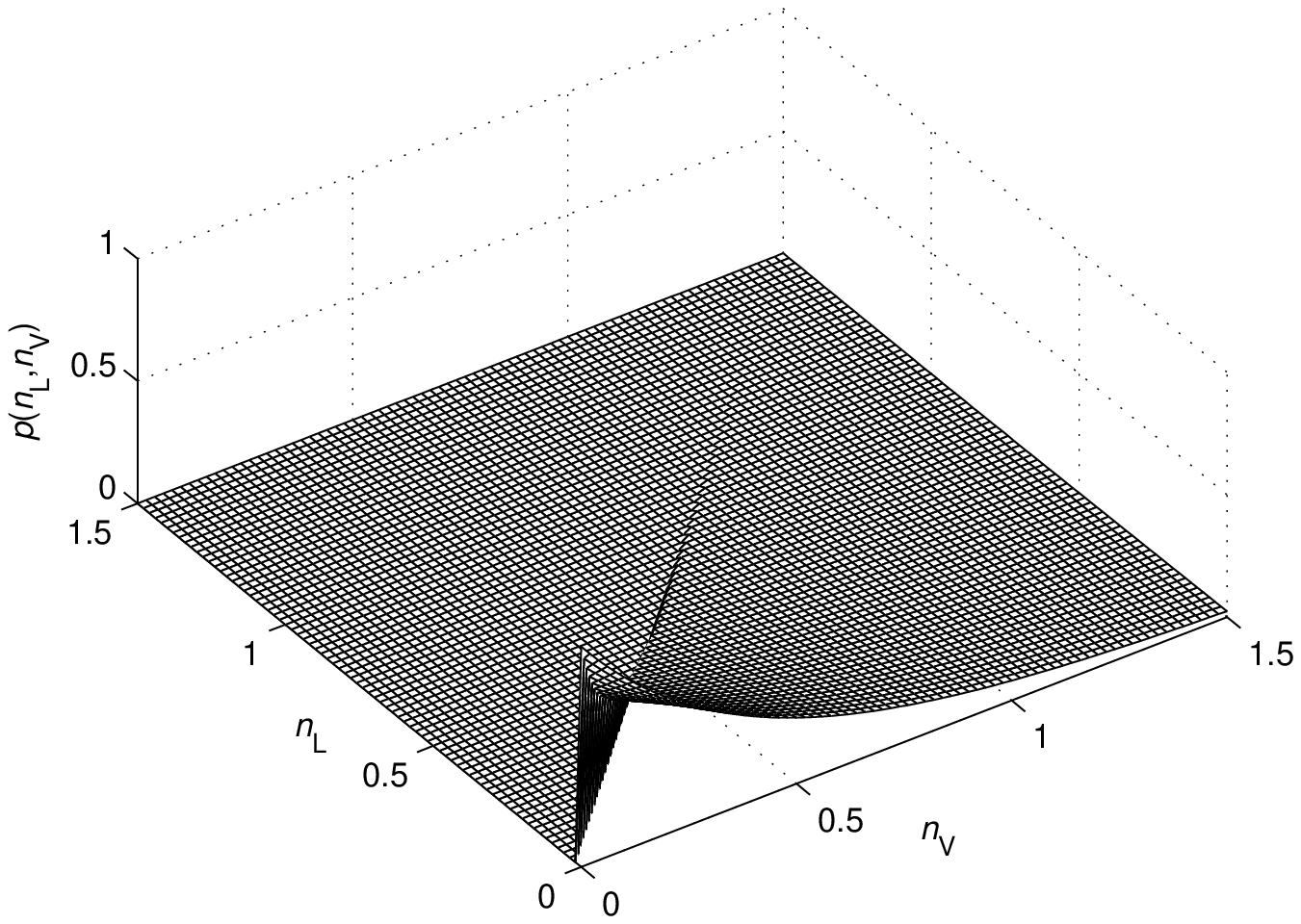}\includegraphics[scale=0.5]{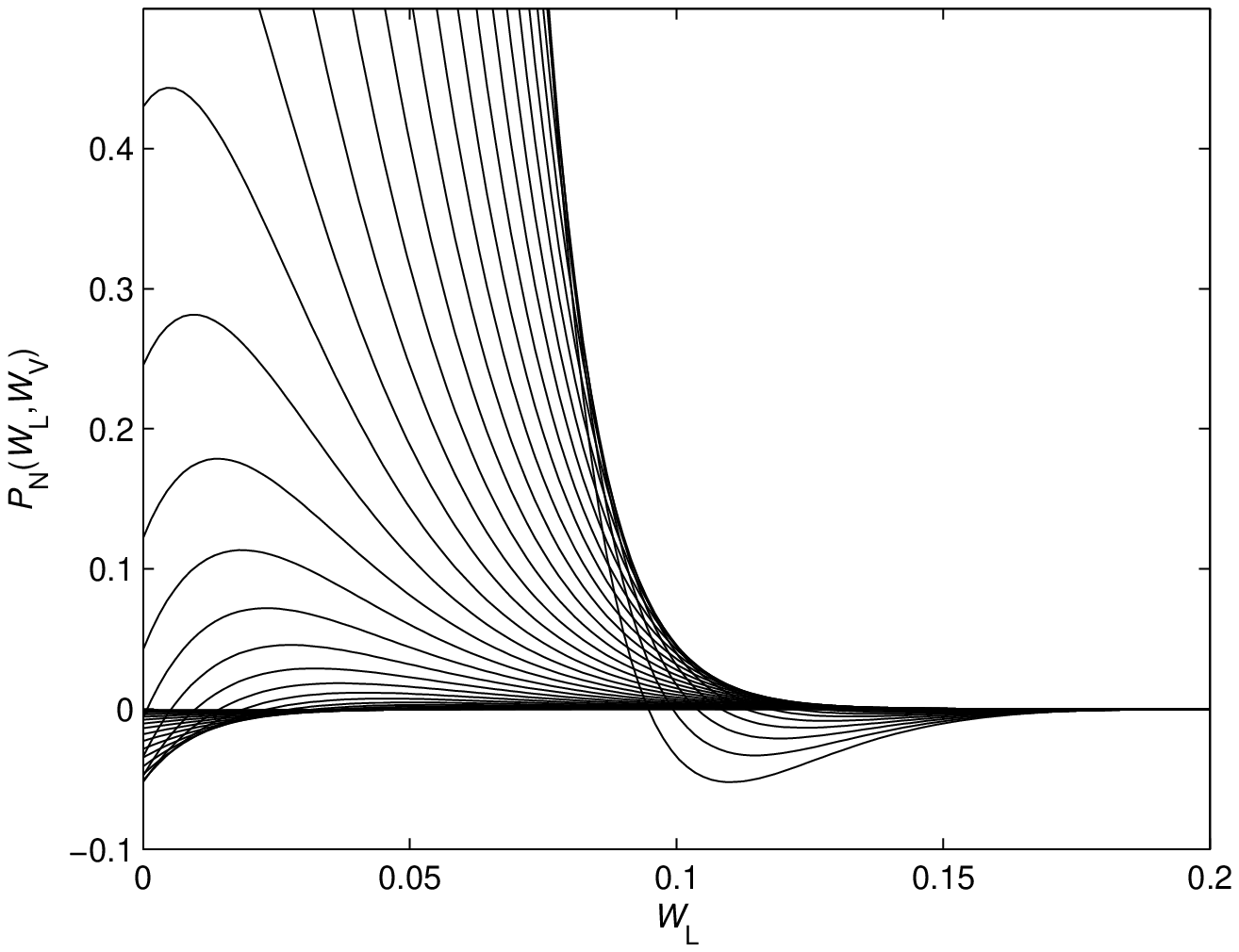}
\par\end{centering}

\caption{\label{fig:pnlnv} a) Joint photon-phonon number distribution for
pump and vibration modes (left) and b) Glauber-Sudarshan quasi-distribution
of integrated-intensities for the same modes with $W_{V}=0,\,0.1,\,0.2.\cdots2$
(right).}
\end{figure}

Similarly, we can treat the compound mode $(LV)$ considering again
photon vacuum scattering with partial stimulation $I_{A}\neq0$ and
$I_{S}=0.$ Shifting distribution in $W_{L}$ along $I_{L}$ and neglecting
short-time terms as above we obtain from (\ref{eq:BCD and Dbar}),
$K_{LV}=-|\chi|^{2}t^{2}I_{A}=-B_{L}$ and $B_{V}=|g|^{2}t^{2}I_{L}+|\chi|^{2}t^{2}I_{A},$
i.e. $K_{LV}+B_{L}=0$ and $K_{LV}+B_{V}=|g|^{2}t^{2}I_{L}>0$ ($C_{L}=C_{V}=\bar{D}_{LV}=0$).
Thus for the joint photon-phonon number distribution \cite{Krepelka-1}
we obtain \begin{equation}
p(n_{L},n_{V})=\frac{n_{V}!}{n_{L}!(n_{V}-n_{L})!}\frac{(B_{V}+K_{LV})^{n_{V}-n_{L}}}{(1+Bn_{V})^{n_{V}+1}},\, n_{V}\geq n_{L}.\label{eq:pustak5}\end{equation}
For $n_{V}<n_{L},$ the distribution is zero. Its quantum behavior
is illustrated in Fig. \ref{fig:pnlnv}a, showing one-side behavior
along the diagonal compared to the earlier cases \cite{Krepelka-2}.
For the threshold values of the ordering parameter we have $s_{th}\approx1+|g|^{2}t^{2}I_{L}-2|\chi|t\sqrt{I_{A}}.$
Assuming for simplicity $|\chi|=|g|,\, I_{A}=1$ and $I_{L}=10,$
we have $s_{th}=0.9$ and nonclassical behavior of wave quasi-distribution
is illustrated by the Glauber-Sudarshan quasi-distribution of integrated-intensities
for $s=1$: \begin{equation}
P_{N}(W_{L},W_{V})=\frac{1}{\pi\sqrt{B_{L}B_{V}}}{\rm e}^{-\frac{W_{L}}{2B_{L}}-\frac{W_{V}}{2B_{V}}}\frac{\sin\left[\frac{\sqrt{\frac{B_{V}}{B_{L}}}W_{L}-\sqrt{\frac{B_{L}}{B_{V}}}W_{V}}{\sqrt{B_{L}}}\right]}{\sqrt{\frac{B_{V}}{B_{L}}}W_{L}-\sqrt{\frac{B_{L}}{B_{V}}}W_{V}},\label{eq:pustak6}\end{equation}
as shown in Fig. \ref{fig:pnlnv}b. The existence of nonclassical
character is clearly visible through the negative values of $P_{N}(W_{L},W_{V}).$

\section{Difference and conditional number distributions \label{sec:Difference-and-conditional}}

\begin{figure}
\begin{centering}
\includegraphics[scale=0.4]{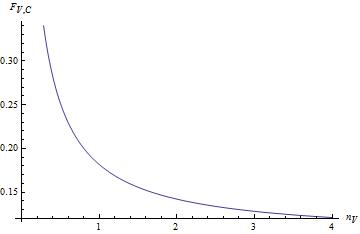}
\par\end{centering}

\caption{\label{fig:Conditional-Fano-factor}Conditional Fano factor $F_{V,C}$
for phonon mode shows nonclassical behavior as $F_{V,C}<1$. Here
$B_{V}=0.11$ and $B_{L}=0.01,$ $K_{LV}=-0.01$.}
\end{figure}

In this section we can further illustrate the observed nonclassicalities
via difference and conditional number distributions. For example,
nonclassical character associated with a mode can be illustrated using
conditional Fano factor, which is defined as \[
F_{i,C}=\frac{\left\langle (\Delta n_{i})^{2}\right\rangle _{C}}{\left\langle n_{i}\right\rangle },\]
for mode $i$. Corresponding condition for nonclassicality is $F_{i,C}<1.$
Analytic expressions for conditional Fano factor are obtained here
for modes of interest (i.e. for $F_{L,C}$ and $F_{V,C}$) as follows:
\begin{equation}
F_{L,C}=1-\frac{B_{L}}{B_{V}},\label{eq:Fano-l}\end{equation}
and \begin{equation}
F_{V,C}=\frac{(n_{L}+1)\left(\frac{1+B_{V}}{1+B_{L}}\right)^{2}-1}{(n_{L}+1)\left(\frac{1+B_{V}}{1+B_{L}}\right)-1}-1.\label{eq:Fano-V}\end{equation}
It is now easy to observe from (\ref{eq:BCD and Dbar}) that $B_{L}$
and $B_{V}$ are always positive, consequently the conditional Fano
factor $F_{L,C}$ is always less than unity. Thus conditional Fano
factor always depicts nonclassicality in pump mode. However, in phonon
mode the presence of nonclassicality (i.e. $F_{V,C}<1$) is not directly
visible from the expression, but the same is shown in the Fig. \ref{fig:Conditional-Fano-factor}.

\begin{figure}
\begin{centering}
\includegraphics[scale=0.4]{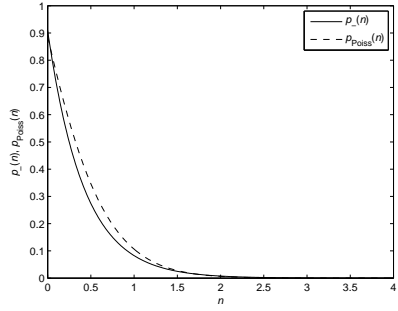}
\par\end{centering}

\caption{\label{fig:Difference-number-distribution}Difference number distribution
$p_{-}(n)$ and $p_{pois}(n)$. Subpoissonian character is shown by
$p_{-}(n)$. Here $B_{V}=0.11$ and $B_{L}=0.01,$ $K_{LV}=-0.01$.}
\end{figure}

The corresponding number distributions are obtained as \begin{equation}
\begin{array}{lcl}
p_{C}\left(n_{L};n_{V}\right) & = & \frac{n_{V}!}{n_{L}!\left(n_{V}-n_{L}\right)!}\left(1-\frac{B_{L}}{B_{V}}\right)^{n_{V}}\left(\frac{B_{L}}{B_{V}-B_{L}}\right)^{n_{L}},\\
p_{C}\left(n_{V};n_{L}\right) & = & \frac{n_{V}!}{n_{L}!\left(n_{V}-n_{L}\right)!}\frac{1+B_{L}}{1+B_{V}}\left(\frac{B_{V}-B_{L}}{1+B_{V}}\right)^{n_{V}}\left(\frac{1+B_{L}}{B_{V}-B_{L}}\right)^{n_{L}}.\end{array}\label{eq:conditional-distribution}\end{equation}
\begin{figure}
\begin{centering}
\includegraphics[scale=0.5]{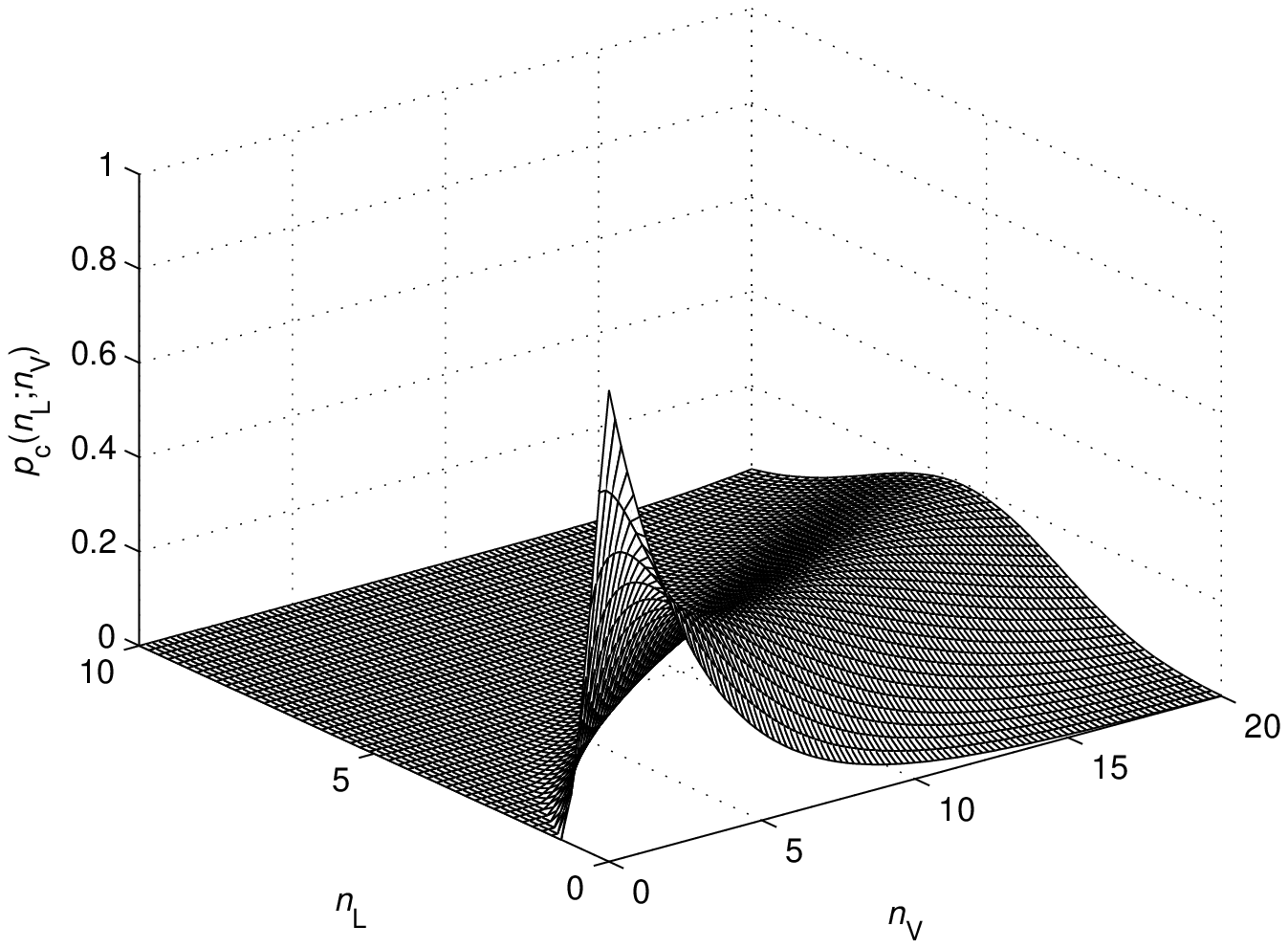}\includegraphics[scale=0.5]{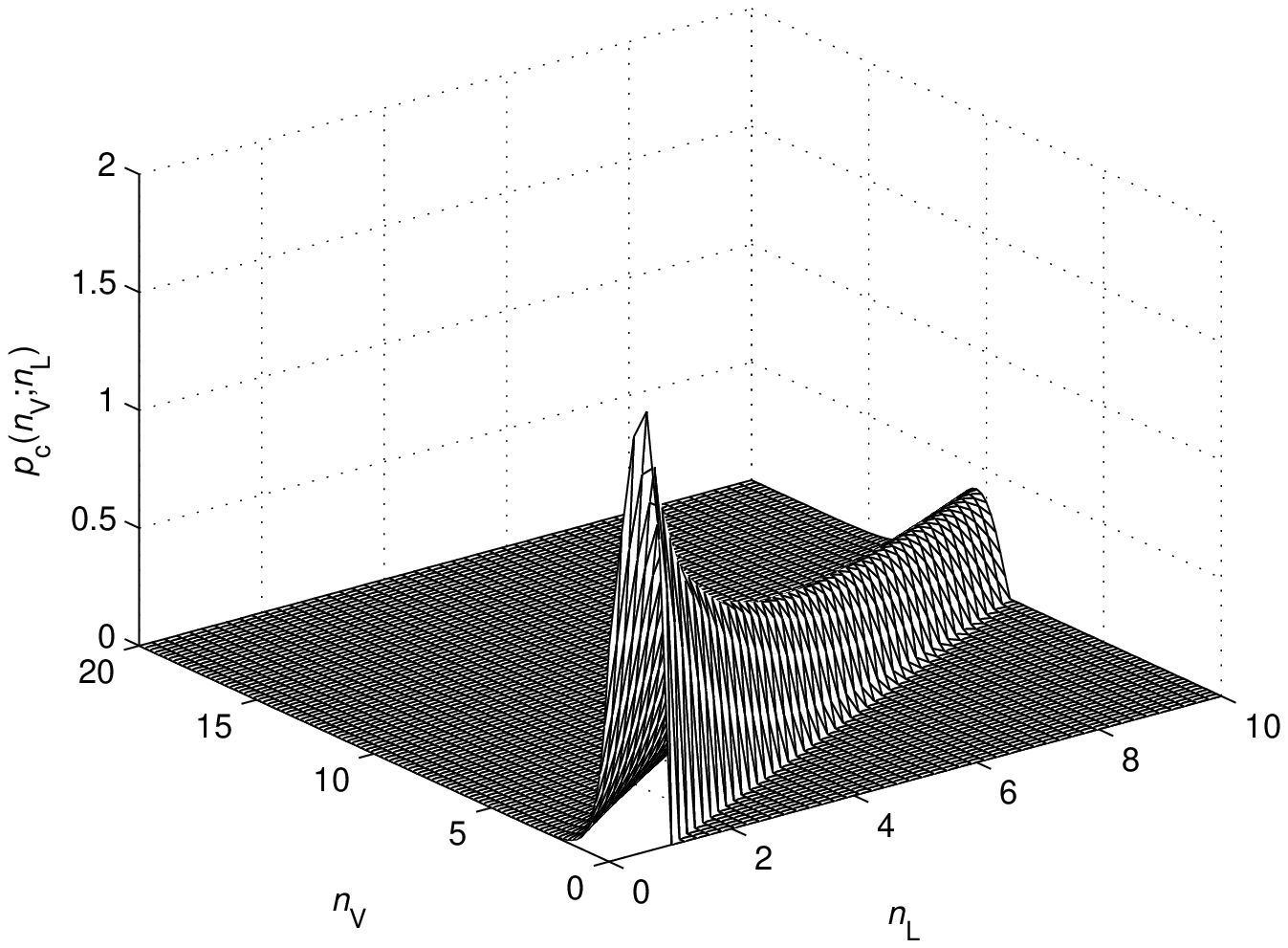}
\par\end{centering}

\caption{\label{fig:Conditional-number-distributions}Conditional number distributions
$p_{C}\left(n_{L};n_{V}\right)$ (left) and $p_{C}\left(n_{V};n_{L}\right)$
(right). Here $B_{V}=0.11$ and $B_{L}=0.01,$ $K_{LV}=-0.01$.}
\end{figure}
These conditional number distributions are plotted in the Fig. \ref{fig:Conditional-number-distributions}.
Difference number distribution can be obtained as \begin{equation}
p_{-}(n)=\frac{\left(B_{V}-B_{L}\right)^{n}}{\left(1+B_{V}-B_{L}\right)^{n+1}},\label{eq:difference-distribn}\end{equation}
\[
\langle\left(\Delta n\right)^{2}\rangle_{-}=\left(B_{V}-B_{L}\right)\left(1+B_{V}-B_{L}\right)\]
and Poissonian distribution for the same two modes is \begin{equation}
p_{Pois}(n)=\frac{\left(B_{V}+B_{L}\right)^{n}}{n!}{\rm e}^{-\left(B_{V}+B_{L}\right).}\label{eq:Poiss-distribn}\end{equation}
A joint plot of $p_{-}(n)$ and $p_{pois}(n)$ is provided in Fig.
\ref{fig:Difference-number-distribution}, which clearly shows subpoissonian
character in $p_{-}(n)$. Thus a nonclassical difference number distribution
is observed. For the sub-shot noise parameter $R=\frac{\langle\left(\Delta n_{ij}\right)^{2}\rangle}{\left(\langle n_{i}\rangle+\langle n_{j}\rangle\right)}$
we have $R\thickapprox1-\frac{2B_{L}}{B_{L}+B_{V}}=0.83<1.$ %

\section{Conclusion\label{sec:Conclusion} }

We have observed different type of nonclassicalities in the stimulated,
completely spontaneous and partially spontaneous Raman process. The
observations that are discussed in detail in Section \ref{sec:Intermodal-entanglement}
are summarized in Table \ref{tab:summary} for coherent scattering.
We see that in general various nonclassical features of the process
can or cannot be directly related, only for combined modes $(LV)$
and $(SV)$ all of them occurs simultaneously. We have not restricted
ourselves to the study of coherent scattering alone. In Section \ref{sec:Phonon-mode-is-chaotic}
we have investigated various nonclassical characters of Raman process
when the phonon mode is chaotic. Finally we have illustrated our results
by joint photon-phonon number and wave distributions. %
\begin{table}[H]
\begin{centering}
\begin{tabular}{|c|>{\centering}p{0.5in}|>{\centering}p{0.8in}|>{\centering}p{0.6in}|>{\centering}p{0.8in}|>{\centering}p{0.8in}|>{\centering}p{0.6in}|>{\centering}p{0.6in}|}
\hline 
Mode & $E_{ij}$ & $\lambda_{ij}$ & $C_{ij}$ & $\left\langle \left(\Delta W_{ij}\right)^{2}\right\rangle _{-}$ & $\left\langle \left(\Delta W_{ij}\right)^{2}\right\rangle _{+}$ & $(K_{ij})_{+}$ & $(K_{ij})_{-}$\tabularnewline
\hline 
AV & non-conclusive & >1 & non-conclusive & +ve & +ve & non-conclusive & non-conclusive\tabularnewline
\hline 
AL & non-conclusive & <1 if $I_{L}>I_{A}$ (expected) & non-conclusive & -ve region exists & -ve region exists & non-conclusive & non-conclusive\tabularnewline
\hline 
AS & non-conclusive & <1 if $|\chi|>|g|$ & non-conclusive & +ve & +ve & non-conclusive & non-conclusive\tabularnewline
\hline 
LS & non-conclusive & >1 & non-conclusive & +ve & -ve region exists & non-conclusive & non-conclusive\tabularnewline
\hline 
LV & always -ve & < for short time & always -ve & -ve region exists & -ve region exists & always -ve & always -ve\tabularnewline
\hline 
SV & always -ve & < for short time & always -ve & -ve & -ve & always -ve & always -ve\tabularnewline
\hline
\end{tabular}
\par\end{centering}

\caption{\label{tab:summary}Negativity of different characteristics of nonclassicality.
Here we have used $E_{ij}=B_{i}B_{j}-|\bar{D}_{ij}|^{2},$ $\lambda_{ij}=1+B_{i}+B_{j}-2{\rm Re}\bar{D}_{ij}-|C_{i}+C_{j}+2D_{ij}|<1,$
$C_{ij}=B_{i}^{2}+B_{j}^{2}+|C_{i}|^{2}+|C_{j}|^{2}-2|D_{ij}|^{2}-2|\bar{D}_{ij}|^{2}<0,$
$\left(K_{ij}\right)_{\pm}=(B_{i}\pm|C_{i}|)(B_{j}\pm|C_{j}|)-\left(|D_{ij}|\mp|\bar{D}_{ij}|\right)^{2}.$}
\end{table}

\textbf{Acknowledgment:} A. P. thanks Department of Science and Technology
(DST), India for support provided through the DST project No. SR/S2/LOP-0012/2010.
He also thanks the Operational Program Education for Competitiveness
- European Social Fund project CZ.1.07/2.3.00/20.0017 of the Ministry
of Education, Youth and Sports of the Czech Republic. J. P. and J.
K. thank the Operational Program Research and Development for Innovations
- European Regional Development Fund project CZ.1.05/2.1.00/03.0058
of the Ministry of Education, Youth and Sports of the Czech Republic.

\end{document}